\documentclass[hyphens
]{ceurart}

\usepackage{listings}
\usepackage{soul}
\newcommand{\multicomment}[1]{}
\newcommand{\dbField}[1]{{\small\texttt{#1}}}

\begin{document}

\copyrightyear{2022}
\copyrightclause{Copyright for this paper by its authors.
  Use permitted under Creative Commons License Attribution 4.0
  International (CC BY 4.0).}

\conference{IAIL'25: Imagining the AI Landscape after the AI Act,
  June 09, 2025, Pisa, Italy}

\title{Improving Regulatory Oversight in Online Content Moderation}

\author[1,2]{Benedetta Tessa}[email=benedetta.tessa@phd.unipi.it
]
\author[3]{Denise Amram}[email=denise.amram@santannapisa.it
]
\author[1]{Anna Monreale}[email=anna.monreale@unipi.it
]
\author[2]{Stefano Cresci}[email= stefano.cresci@iit.cnr.it
]
\address[1]{University of Pisa, Pisa, Italy}
\address[2]{IIT-CNR, Pisa, Italy}
\address[3]{Scuola Superiore Sant'Anna, Pisa, Italy}

\begin{abstract}
The European Union introduced the Digital Services Act (DSA) to address the risks associated with digital platforms and promote a safer online environment. However, despite the potential of components such as the Transparency Database, Transparency Reports, and Article 40 of the DSA to improve platform transparency, significant challenges remain. These include data inconsistencies and a lack of detailed information, which hinder transparency in content moderation practices. Additionally, the absence of standardized reporting structures makes cross-platform comparisons and broader analyses difficult. To address these issues, we propose two complementary processes: a \textit{Transparency Report Cross-Checking Process} and a \textit{Verification Process}. Their goal is to provide both internal and external validation by detecting possible inconsistencies between self-reported and actual platform data, assessing compliance levels, and ultimately enhancing transparency while improving the overall effectiveness of the DSA in ensuring accountability in content moderation.  Additionally, these processes can benefit policymakers by providing more accurate data for decision-making, independent researchers with trustworthy analysis, and platforms by offering a method for self-assessment and improving compliance and reporting practices.

\end{abstract}


\maketitle

\section{Transparency in Online Content Moderation} The Digital Services Act (DSA) is a regulatory framework introduced by the European Union (EU) in 2022 to enhance transparency, accountability, and user protection in digital services~\cite{eu2020DSA}. It was designed to address growing concerns related to the power and influence of large online platforms, particularly in tackling illegal content, disinformation, and systemic risks. By establishing clear obligations for online platforms, the DSA aims to create a safer and more transparent digital ecosystem.

A key element of this framework is the tool adopted by the EU Commission for transparency purposes, namely the Transparency Database (\texttt{DSA-TDB}),\footnote{\url{https://transparency.dsa.ec.europa.eu/}} launched in September 2023 with the goal to enhance transparency and accountability in digital content governance. It is a centralized public repository where Very Large Online Platforms (VLOPs) and Very Large Online Search Engines (VLOSEs) are required to submit structured, detailed, and timely information about each moderation action taken in the EU. Each action is reported to the \texttt{DSA-TDB} by submitting a \textit{Statement of Reasons}\footnote{\url{https://www.eu-digital-services-act.com/Digital_Services_Act_Article_17.html}} (SoRs)---a database record that provides detailed information on the moderation action. SoRs are composed of multiple attributes, including required, conditionally required, and optional ones. They can be filled either via predefined sets of categories or via free text. These statements are meant to clearly explain why content has been removed, restricted, or otherwise affected by a platform’s moderation actions. By requiring platforms to disclose reasons for their moderation actions, the \texttt{DSA-TDB} is aimed at promoting transparency in content governance, thus enabling users, researchers, and policymakers to better scrutinize the enforcement policies of online platforms \cite{trujillo2023dsa,kaushal2024automated,papaevangelou2024content}.

In addition to the \texttt{DSA-TDB}, the Transparency Reports\footnote{\url{https://www.eu-digital-services-act.com/Digital_Services_Act_Article_15.html}} are another component required by the DSA. These reports are thought to provide detailed insights into how platforms enforce their moderation policies, and they must be clear and easily accessible \cite{urman2023transparent}. While the \texttt{DSA-TDB} provides granular, moderation action-level data daily, the Transparency Reports deliver aggregated, periodic insights into how platforms enforce their moderation policies over time. These reports, which must be published at least once a year, include statistics on content removals, the use of automated systems, and efforts to tackle disinformation and other systemic risks. Their purpose is to offer a broader overview of content moderation practices in a format that is more accessible and straightforward for regulators, researchers, and the general public.
Finally, to strengthen transparency and oversight, Article 40\footnote{\url{https://www.eu-digital-services-act.com/Digital_Services_Act_Article_40.html}} introduces a mechanism that is intended to enable regulatory authorities and vetted researchers to access platform data under specific conditions. This article was designed to support independent assessments of platform compliance with DSA obligations and to facilitate research aimed at identifying and mitigating systemic risks in the EU, such as the spread of disinformation, illegal content, and other phenomena with significant societal impact.
Despite its goals, accessing data is not easy \cite{jaursch2024enabling, goanta2025great}.
In practice, researchers can submit a request outlining the aims of their project, the data needed, how long and in which way data will be stored, and how the proposed project assesses the systemic risks. Each request undergoes an evaluation process that determines whether it meets the criteria established in Article 40. Some criteria include being affiliated with a recognised research organization, acting independently from any commercial interests, and most importantly, contributing to the detection, identification, and understanding of systemic risks in the EU. Only approved requests may result in formal data access being granted by the platform. Depending on the sensitivity of the data, access may take the form of direct transmission or, for more sensitive cases, use of secure processing environments.
A delegated act on data access is currently under consultation to clarify the procedures for requesting and accessing platform data, ensuring effective oversight across platforms\footnote{\url{https://ec.europa.eu/info/law/better-regulation/have-your-say/initiatives/13817-Delegated-Regulation-on-data-access-provided-for-in-the-Digital-Services-Act_en}}.

Ultimately, while the DSA introduces the aforementioned components to enhance transparency and accountability, its effectiveness depends on how platforms leverage them and what they choose to report.

 \section{Transparency and Accountability Challenges}
Since the introduction of the DSA, several studies have assessed the effectiveness of its key transparency components, namely the \texttt{DSA-TDB}, the Transparency Reports, and the obligations described in Article 40. As a result, many critical issues have emerged, particularly regarding data inconsistencies, lack of detail, and lack of transparency.

A key issue concerns the limited informativeness of the records submitted to the \texttt{DSA-TDB} due to the frequent use of overly generic categories in required attributes. In fact, platforms tend to vaguely attribute moderation decisions to a violation of their term of service without going into details on which term was violated and why \cite{shahi2025year,trujillo2023dsa}.
Moreover, this vagueness and lack of detail is exacerbated by an overall underutilization of optional-yet potentially insightful—attributes \cite{trujillo2023dsa,kaushal2024automated}. For example, the \dbField{decision\_ground\_reference\_url} attribute, which should provide users with direct access to the legal or contractual basis of a moderation decision, is rarely filled. As a result, users are left without clear references or understanding of platform actions. Similarly, the \dbField{illegal\_content\_explanation} attribute is largely omitted, meaning that when content is flagged as illegal, platforms often fail to specify why, making it difficult to assess the legitimacy of these decisions. Moreover, a misuse of certain attributes has been found. For example, Snapchat recorded notifications from ``Trusted Flaggers'' in the database before this role was officially established at the EU level, seemingly conflating the DSA framework with an internal initiative of the same name \cite{trujillo2023dsa,dergacheva2023one}.
Another critical issue is the structure of the database itself, which fails to facilitate transparency and does not meet the reporting needs of platforms~\cite{trujillo2023dsa}. For instance, there is no dedicated field to indicate whether a moderation action targets an account rather than specific content, forcing platforms to classify such cases under \texttt{other} and clarify in free-text fields. 

This misuse of the database reflects a broader issue, as many online platforms report inconsistent data. A notable example is X (formerly Twitter), which exhibited the highest number of inconsistencies, particularly regarding automated moderation. Despite Transparency Reports indicating some level of automation, the platform reported no automated actions in the \texttt{DSA-TDB} \cite{trujillo2023dsa,kaushal2024automated,dergacheva2023one}. This discrepancy was also observed for Facebook and Instagram, that reported extensive use of fully automated moderation in their transparency reports but did not indicate any fully automated actions in the \texttt{DSA-TDB} \cite{trujillo2023dsa}.
Beyond data inconsistencies, other issues have emerged in Transparency Reports. Many platforms focus on aggregate data without details on specific decisions, present their policies ambiguously, and prioritize government requests over their own moderation actions. In particular, they also often omit key information, such as whether users can appeal moderation decisions, how platforms handle misuse of reporting tools, and whether content removal requests come from private entities, like copyright holders, rather than only from governments \cite{urman2023transparent}. In general, many platforms fail to provide clear methodologies for compiling the statistics presented in their Transparency Reports. Additionally, the reports are often inconsistent in their structure, making it difficult to compare data across platforms or aggregate it for a broader analysis \cite{bommasani2024foundation}. Additionally, the Oversight Board\footnote{\url{https://www.oversightboard.com/news/statements-from-the-oversight-board-trust-and-oversight-board-members-on-the-announcement-of-the-appeals-centre-europe/}}—an independent body created by Meta to review and make binding decisions on content moderation cases—has pointed out that Transparency Reports often underrepresent interactions with law enforcement, raising concerns about potential biases in the flagging process and the lack of opportunities for users to contest or respond to flagged content \cite{van2025article}. Overall, these issues arise not only from the self-reported nature of the information required by the DSA, which allows a selective disclosure of information, but also from the structure of the reporting tools, which enables platforms to meet the formal compliance requirements of the DSA while providing only minimal or vague details. This significantly limits the transparency and informativeness that the \texttt{DSA-TDB} was designed to promote \cite{kaushal2024automated}. Moreover, significant discrepancies in moderation practices and the application of DSA obligations across platforms result in a lack of shared standards, curbing the DSA's goal of standardization across all digital platforms \cite{drolsbach2024content}. In addition to these 
limitations, several challenges have also arisen with the implementation of Article 40 of the DSA, particularly for researchers. Request forms are often difficult to find and are unclear, while eligibility criteria tend to be excessively restrictive. Responses from platforms are typically slow and vague, and the available data lacks sufficient documentation. To better support independent research, more precise guidelines, broader eligibility, faster responses, and continued access to data are essential \cite{jaursch2024enabling}.

Due to these compliance issues, some platforms are facing formal proceedings initiated by the European Commission under the DSA. For X\footnote{\url{https://ec.europa.eu/commission/presscorner/detail/en/ip_23_6709}}, the proceedings focused on failures related to countering illegal content, the effectiveness of its notice-and-action system, and shortcomings in transparency, particularly in providing researchers with access to public data as required by Article 40. Other concerns include the platform's risk assessment processes and design choices, such as using checkmarks linked to subscription products. Similarly, TikTok\footnote{\url{https://ec.europa.eu/commission/presscorner/detail/en/ip_24_2227}} is under investigation for not conducting proper risk assessments before rolling out features like the ``Task and Reward Lite'' program, which may harm minors' mental health. Meta's\footnote{\url{https://ec.europa.eu/commission/presscorner/detail/en/ip_24_2664}} proceedings concern its failure to adequately address risks associated with Facebook and Instagram interface designs, which may exploit minors and cause addictive behavior. 

To conclude, the \texttt{DSA-TDB}, Transparency Reports, and Article 40 each offer valuable instruments to enhance transparency in content moderation. However, in practice, all three have faced significant limitations in terms of data quality, consistency, and accessibility, which have hindered their ability to meet the DSA’s transparency goals. For example, it has been demonstrated that, in its current state, the SoRs stored in the \texttt{DSA-TDB} fail to answer simple research questions that should instead be addressable \cite{shahi2025year}. This gap highlights the need for further improvements to ensure that the tools at hand can effectively support transparency and accountability in content moderation. Nonetheless, the fact that these tools provide different angles on the same problem suggests that combining them could help address some of their weaknesses and lead to a more meaningful understanding of platforms' content moderation decisions.
 \begin{figure}[t!]
    \centering
    \includegraphics[width=\linewidth]{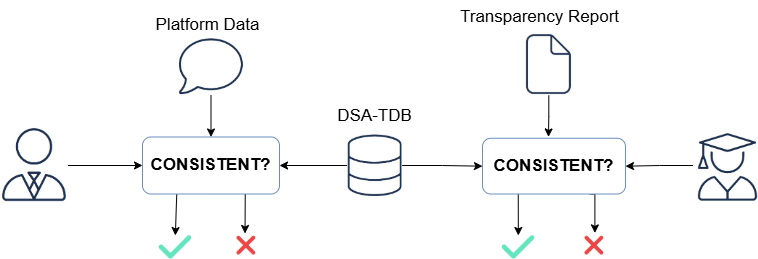}
    \caption{Overview of the proposed complementary processes for enhancing transparency and detecting inconsistencies in platforms' reported moderation action. The consistency check between the SoRs in the \texttt{DSA-TDB} and Transparency Reports verifies coherency, while the check between the SoRs and platform shared data verifies their completeness and trustworthiness.}
    \label{fig:process}
\end{figure}

\section{Towards Greater Transparency and Legal Accountability}

\subsection{Verification Processes for Transparency Assessment}

While the \texttt{DSA-TDB}, the Transparency Reports, and Article 40 of the DSA hold the potential to advance platform transparency, their effectiveness and reliability are limited by the absence of verification mechanisms. 
To address these issues, we propose two complementary processes designed to enhance transparency and detect inconsistencies in platforms' reports, as shown in Figure \ref{fig:process}. 

\subsubsection{Transparency Report Cross-Checking Process}
Ensuring consistency between Transparency Reports and the \texttt{DSA-TDB} is crucial for evaluating whether platforms provide truthful reports of their enforcement practices. Transparency Reports typically present aggregate statistics over a certain period of time, whereas the \texttt{DSA-TDB} collects structured records of individual enforcement actions. While the \texttt{DSA-TDB} and Transparency Reports are complementary, current research has highlighted various mismatches between them. To address this issue, we propose a \textit{Transparency Report Cross-Checking Process} that performs an internal validation by comparing the data reported in Transparency Reports with the corresponding aggregations derived from the \texttt{DSA-TDB}. 

Firstly, the process would start by extracting and identifying all of the aggregations presented in the Transparency Reports. However, they are typically published in unstructured formats, such as PDFs or dedicated web pages, and rarely offer access to raw, downloadable data. Moderation statistics are often embedded in visual formats, like charts or tables, which makes automated extraction challenging. In this case, the process should include a preprocessing module that uses, for instance, optical character recognition (OCR) and rule-based parsing to extract data from PDF tables and charts. Large language models (LLMs) and other automated tools have also proven to be effective in extracting data from PDF documents \cite{li2024extracting,khraisha2024can,zhu2022pdfdataextractor}. Additionally, HTML parsing can be used to retrieve aggregates from web-based reports. Once the target aggregations are extracted, the process will automatically replicate them by using the SoRs from the \texttt{DSA-TDB}. SoRs are freely downloadable and include attributes like violation type and date, allowing for filtering and matching each aggregation in the Transparency Reports. To give an example, if a platform reports the removal of $N$ comments for hate speech within a specific time frame, the process can retrieve the corresponding SoRs from the \texttt{DSA-TDB} for the same period, content type, and violation category, then perform the necessary aggregations based on the data in the Transparency Reports. Finally, the aggregated results from both sources would be compared to assess consistency. If any discrepancies are found, these can be flagged for further investigation.

Since both the Transparency Reports and the \texttt{DSA-TDB} are self-reported, the goal is not to assess their objective accuracy but to verify their internal coherence.  As a result, the process can not only support platforms in verifying the coherence of their own reports before publication, but also assist regulators, researchers, and external auditors in evaluating the reliability of the self-reported data.

\subsubsection{Verification Process}
As discussed in the previous sections, while the \texttt{DSA-TDB} provides valuable insights into individual moderation actions taken by platforms, its reliability has raised several concerns. Platforms retain full control over what information to include and how to describe it, which can lead to selective reporting, misclassification, or omission of moderation actions.
To verify the trustworthiness and the completeness of the SoRs submitted to the \texttt{DSA-TDB}, a promising approach involves comparing them with the actual platform data. As a matter of fact, Article 40 of the DSA allows competent authorities and vetted researchers to request this original content and metadata, possibly enabling external validation of the actions reported in the \texttt{DSA-TDB}.
For this reason, we propose a \textit{Verification Process} that serves as an external assessment mechanism to compare the SoRs in the \texttt{DSA-TDB} with the platform data.
This process can be used by regulators to support audits, verify the truthfulness of the information submitted to the \texttt{DSA-TDB}, and identify discrepancies in reported enforcement actions. Moreover, digital service providers themselves may use it internally as a way to reassess risks and ensure that the information they report is consistent with their actual moderation practices. Similarly, bodies responsible for codes of conduct could rely on this process to evaluate the level of compliance of signatories and improve coordination across members. Over time, this could even support the development of shared certification standards, helping define what counts as trustworthy reporting under the DSA framework.
To streamline this process, given the large volume of data involved, Artificial Intelligence (AI) can play a key role. AI tools may offer significant advantages by enabling a quick analysis of large datasets and efficiently identifying the types of moderated content. For example, these systems can detect various forms of online harm subject to platform moderation, such as toxicity and hate speech \cite{cima2025contextualized,mullah2021advances}, misinformation \cite{abdali2024multi}, deepfakes \cite{almars2021deepfakes}, and nudity \cite{tabone2021porn}. 

More in detail, the \textit {Verification Process} could operate by first extracting all moderated content (e.g., comments, videos, or images) within a specified time frame from the platform data. Both the moderation status and whether this content was posted within a given time window can be retrieved from the content metadata, such as visibility status, assigned violation categories, and the original creation date of the content. Moreover, AI algorithms would be leveraged to automatically classify this content based on different violation types, such as hate speech, misinformation, or nudity. This classification can be implemented through a combination of natural language processing and image recognition technologies, ensuring a multimodal approach to identifying different forms of harmful content. The process would then proceed to extract additional information from the content metadata, such as timestamps, content type, and any platform-specific flags or annotations. This information is essential for reconstructing the corresponding Statements of Reasons (SoRs) for each piece of content, according to the guidelines set by the DSA. These reconstructed SoRs would then be compared with the original SoRs stored in the DSA-TDB from the same time period to identify any potential discrepancies. If any are found, the process will flag them for further review, providing authorities with the information necessary to ensure transparency and accountability in content moderation processes.

\subsection{Legal Implications}
The DSA has introduced due diligence obligations including the risks detection and evaluation in order to ensure a safer access to digital services. However, it lacks in identifying specific technical and organizational safeguards to prevent and mitigate the detected risks, requiring each platform to identify the proper measure according to the characteristics of the service and their users. 
In this regard, the way platforms implement transparency obligations, such as providing SoRs and providing access to aggregate data in the Transparency Reports, has a significant impact on the DSA enforcement and effectiveness.
Firstly, users might be more aware about unlawful conducts and avoid to perpetrate them, while digital services providers that are not meeting the requirements of ``very large'' ones, may in any case be supported in their assessments and evaluations in order to prevent and mitigate their corresponding risks of sharing unlawful, discriminatory, or harmful contents. 
Furthermore, also considering the huge number of data available, specific analyses could either provide tailored variables to suggest law and policy makers recommendations or shape a paradigm of accountability for economic operators that could be useful for self-regulatory process (like codes of conducts) to promote a safer and trustworthy digital environment.
In fact, the opportunity to directly compare scenarios of unlawful use of the digital environment, facilitated by processes like the \textit{Transparency Report Cross-Checking Process} and the \textit{Verification Process}, helps identify trends and improve the system by providing harmonized solutions for managing and reporting unlawful behaviors online.
These premises strongly confirm whether data included in the Transparency Reports and in the \texttt{DSA-TDB} are accurate and homogeneous. Otherwise, both the achievement of deterrence purposes and the enforcement ones can be dramatically compromised. 
From this perspective, the monitoring mechanisms of the compliance life-cycle under the DSA might be supported by technical process able to provide consistency checks between the aggregate information emerging by the Transparency Platforms, the report obligations, and the real time analyses of the digital environment. 
These processes might also play an effective role in terms of certification and standardisation of the level of compliance. 
 \section{Conclusions}
In this work, we debated the level of transparency provided by the DSA with respect to online content moderation, with a focus on the \texttt{DSA-TDB}, Transparency Reports, and Article 40. Recent literature has highlighted that, while unprecedented, these components in their current form do not fully meet the transparency and accountability objectives they were designed for due to limitations in data quality, access, and reliability.
To truly achieve these objectives, an additional layer of verification and validation is required to address the existing flaws in the components, ensure that the data shared by platforms reflects their actual enforcement actions, and provide scrutiny on self-reported data.
Without mechanisms to verify and cross-check the reported data, regulators, researchers, and the general public cannot fully rely on these reports to understand platform behavior and evaluate the effectiveness of moderation practices.

To this end, we proposed two processes: the \textit{Transparency Report Cross-Checking Process}, whose goal is to internally validate the coherency between Transparency Reports and the \texttt{DSA-TDB} records, and the \textit{Verification Process}, which allows external validation by comparing the platform shared data with the SoRs in the \texttt{DSA-TDB}. Although one process focuses on internal and the other on external validation, they are complementary and, when used together, would offer their maximum potential. These processes are intended to fully leverage the unprecedented components of the DSA, helping to ensure that its promises of transparency and accountability are effectively fulfilled. In addition, they may also contribute to the development of certification and standardisation practices, supporting a more reliable and systematic approach to compliance. This represents a promising direction for future research, as it can support the evolution of digital regulations and contribute to the development of more effective transparency and accountability mechanisms.
 
\section{Acknowledgments}
This work is partly supported by the ERC project DEDUCE under grant \#101113826, the European Union -- Next Generation EU, Mission 4 Component 1, for project PIANO (CUP B53D23013290006), the European Commission under the NextGeneration EU programme – National Recovery and Resilience Plan (PNRR), under agreements: PNRR - M4C2 - Investimento 1.3, Partenariato Esteso PE00000013 - ``FAIR - Future Artificial Intelligence Research" - Spoke 1 "Human-centered AI", and SoBigData.it – ``Strengthening the Italian RI for Social Mining and Big Data Analytics” – Prot. IR0000013 – Avviso n. 3264 del 28/12/2021, SoBigData RI PPP GA 101079043.

\multicomment{
\begin{acknowledgments}
  Thanks to the developers of ACM consolidated LaTeX styles
  \url{https://github.com/borisveytsman/acmart} and to the developers
  of Elsevier updated \LaTeX{} templates
  \url{https://www.ctan.org/tex-archive/macros/latex/contrib/els-cas-templates}.  
\end{acknowledgments}
}
\multicomment{
\section*{Declaration on Generative AI}

 During the preparation of this work, the authors used ChatGPT in order to: Grammar and spelling check. After using this tool, the authors reviewed and edited the content as needed and take full responsibility for the publication’s content. 
}
\bibliography{bibliography}

\appendix

\end{document}